\documentclass[prb,showpacs,superscriptaddress,twocolumn]{revtex4}
\usepackage{eurosym}
\usepackage{amsmath}
\usepackage{amssymb}
\usepackage{amsfonts}
\usepackage{graphicx,hyperref}
\usepackage{dcolumn}
\usepackage{bm}
\usepackage{color}
\begin{document}

\title{Optimal control of hybrid qubits: implementing the quantum permutation algorithm}

\author{C. M. Rivera-Ruiz}
\affiliation{Departamento de F\'{\i}sica, Universidade Federal de S\~ao Carlos, 13565-905, S\~ao Carlos, SP, Brazil}
 
\author{E. F. de Lima}
\affiliation{Departamento de F\'{\i}sica, Universidade Federal de S\~ao Carlos, 13565-905, S\~ao Carlos, SP, Brazil}

\author{F. F. Fanchini}
\affiliation{Faculdade de Ci\^encias, UNESP - Universidade Estadual Paulista, Bauru, SP, 17033-360, Brazil}

\author{V. Lopez-Richard}
\affiliation{Departamento de F\'{\i}sica, Universidade Federal de S\~ao Carlos, 13565-905, S\~ao Carlos, SP, Brazil}

\author{L. K. Castelano}
\email{lkcastelano@ufscar.br}
\affiliation{Departamento de F\'{\i}sica, Universidade Federal de S\~ao Carlos, 13565-905, S\~ao Carlos, SP, Brazil}

\date{\today}

\begin{abstract}
The optimal quantum control theory is employed to determine electric pulses capable of producing quantum gates with high fidelity (higher than $0.9997$). Particularly, these quantum gates were chosen to perform the permutation algorithm (Z. Gedik \textit{et al.}, Scientific reports \textbf{5}, 14671, (2015).) in hybrid qubits in double quantum dots (DQDs). The permutation algorithm is an oracle based quantum algorithm that solves the problem of the permutation parity faster than a classical algorithm without the necessity of entanglement between particles. The only requirement for achieving the speedup is the use of a one-particle quantum system with at least three levels. The high fidelity found in our results is closely related to quantum speed limit, which is a measure of how fast a quantum state can be manipulated. Furthermore, our scheme can be used for the practical realization of different quantum algorithms in DQDs.
\end{abstract}

\newpage
%\pacs{03.65.Ta, 03.67.Mn, 42.50.Dv}

%\pacs{03.65.-w; 02.20.-a; 21.10.sf; 31.30.jx}
%\keywords{}

\maketitle
%\end{CJK*}

\section{Introduction}
During the last decades there has been an active interest in the application of quantum phenomena to outperform classical devices in the same computational task, specially since the proposal of the Grover~\cite{Grover} and Shor~\cite{Shor} algorithms. Quantum protocols exploit resources like the entanglement or the quantum superposition to achieve the speedup on certain tasks, \textit{e.g.} the quantum cryptography protocol BB84 in Ref.~\onlinecite{BB84}. Aiming at experimentally creating such quantum protocols, different platforms have been studied, such as quantum electrodynamic cavities \cite{Cavity}, superconducting circuits \cite{supercondutor}, and ion traps in ultra-cold atoms \cite{iontrap}.
Some platforms, such as Rydberg atoms~\cite{Rydberg1}, nitrogen-vacancy centers~\cite{Vacancy}, and quantum dots (QDs)~ \cite{DiVincenzo},
employ the electronic spin, which is relatively more isolated
from the surrounding environment than other systems.
In particular, semiconductor QDs is a promissing platform, where the required scalability to perform large-scale quantum computation might be accomplished~\cite{QDadvant3}. The information can be encoded either in the electronic spin (known as spin qubits) or in the discrete electronic levels due to the spatial confinment (known as charge qubits). The implementation of single and two qubit gates by means of spin qubits paves the way to the realization of universal quantum computing \cite{Dloss}. %pag 23
The main strength of this platform comes from the fact that spins have longer lifetimes compared to charges, although spins require longer manipulation times. Devices based on charge qubits have manipulation times of the order of picoseconds, thus providing fast information processing \cite{Shi2014, Dloss}. A new platform called hybrid qubit emerged a few years ago~\cite{hybrid1}, which combine both spin and charge qubits in DQDs to achieve fast manipulation and long decoherence time~\cite{hybrid2,hybrid3,hybrid4,hybrid5,hybrid6}.

The precise control of quantum phenomena is a long-standing dream, specially when it is related to the implementation of quantum information protocols. Optimal quantum control theory offers the possibility to control the physical system by maximizing a certain observable and has succeeded in several platforms. For example, in spin qubits of nitrogen-vacancy centers in Ref.~\onlinecite{connitro}, there is a remarkable difference in the fidelities of the output states when optimal pulses are used compared to the standard pulses. In the Bose-Einstein condensate~\cite{conBose}, the error in the preparation of a state is dramatically reduced by using optimal control. There are also other examples in the superconducting circuits~\cite{consuperconduct1,consuperconduct2}, where a CNOT gate was realized with success.

In the present study, we apply the optimal quantum control theory in hybrid qubits platform. Our goal is to demonstrate the possibility of electrically implementing quantum gates with high fidelity in such systems. As an example, we choose the quantum permutation algorithm (QPA) Ref.~\onlinecite{Gedik2015}, which requires quantum superposition of states with well-defined relative phases. 
Because of the necessity of using at least a three level system in this algorithm, we use hybrid qubits instead of spin qubits.
In order to find the optimal AC electric fields that implement the required quantum gates, we apply a technique known as the two-point boundary-value quantum control paradigm (TBQCP)~\cite{control2010}. By employing such a technique, we are able to determine optimal electric pulses that perform the quantum gates with high fidelity and in times faster than the decoherence time. Our results open the possibility of achieving all-electrical universal quantum gates in DQDs by means of optimal quantum control.

\section{Theoretical Framework}
\subsection{Optimal Quantum Control}
In order to implement quantum gates with high fidelity, we use the optimal quantum control theory. The basic goal of optimal control consists of finding time-dependent control fields that drives an initial state $\mid \psi(0) \rangle$ to a specific target state $\mid \psi_{tar} \rangle$ at the end of the time evolution. In this study, we implement the numerical method TBQCP~\cite{control2010}, which is an iterative monotonic method able to find an optimal field $E_{opt}(t)$ that maximizes the expectation value of a physical observable $\langle O(t)\rangle$ at the final time $T$. This method starts with the definition of the boundary conditions, the initial state $\mid \psi(0) \rangle$ and the desired physical observable $\langle O(T)\rangle$. The physical observable is evolved backwards (from the final time $T$ to the initial time $t=0$) through the following equation 

\begin{equation}\label{oper}
i\text{\ensuremath{\hbar}} \frac{\partial O^{(n)}(t)}{\partial t}=\left[O^{(n)}(t),\, H_{0}-\mu E^{(n)}(t)\right],\; O(T) \rightarrow O(0),
\end{equation}
where $H_{0}$ is the time independent Hamiltonian of the system, $\mu$ is the dipole operator, and $E^{(n)}(t)$ is the field in the nth iteration. The initial state $\mid \psi(0) \rangle$ is evolved forward with the time-dependent Schr\"odinger equation, \begin{equation}\label{SEqu}
 i\text{\ensuremath{\hbar}}\frac{\partial\mid{\psi^{(n+1)}(t)}\rangle}{\partial t}=\left( H_{0}-\mu E^{(n+1)}(t)\right) \mid\psi^{(n+1)}(t)\rangle,%\;\; \mid{\psi(0)}\rangle \rightarrow \mid{\psi(T)}\rangle,
 \end{equation} 
where $E^{(n+1)}(t)$ is the (n+1)st iteration field, which is calculated through the following expression
\begin{equation}
E^{(n+1)}(t)=E^{(n)}(t)+ \eta f^{(n+1)}_{\mu}(t).\label{fieldn}
\end{equation}
In Eq.~(\ref{fieldn}), $\eta$ is a positive constant and the field correction is written as
\begin{equation}
f^{(n+1)}_{\mu}(t) = -\frac{2}{\hbar}\textrm{Im}\left\{ \langle \psi^{(n+1)}(t)| O^{(n)}(t) \mu  \mid \psi^{(n+1)}(t)\rangle \right\},\label{fmu}
\end{equation}
Equations~(\ref{oper}-\ref{fmu}) are solved in a self-consistent way, starting with the trial field $E^{(0)}(t)$ and monotonically increasing the value of the desired physical observable $\langle O(T)\rangle=\langle\psi(T)\mid O(T)\mid\psi(T)\rangle$, see more detaisl in Ref.~\onlinecite{control2010}. Particularly, if one is interested in maximizing an specific target state $\mid\psi_{target}\rangle$, the observable becomes the projector onto this state $\langle O(T)\rangle=|\langle\psi(T)\mid\psi_{target}\rangle|^2$. 
%To implement quantum gates, we need to construct an optimized field that acts on each state of the logical basis plus a particular linear combination of all states of the logical basis aimig at performing the truth table. The last constraint imposed to the optimized field is necessary to avoid relative phase errors. Such scheme of implementing quantum gates has already been been discussed in Refs.~\onlinecite{multitarget,Kosloff}.

\subsection{Implementation of Quantum Gates}
Quantum gates are the quantum analogue of logic gates in classical computers. Such gates are reversible in time and are represented by unitary matrices $\mathcal{U}$. The optimal quantum control scheme also can be employed to implement quantum gates~\cite{universal,Kosloff}. Such an implementation can be performed by finding the optimal field that guides a set of initial $k$-states $\mid \psi_{i,k}(0)\rangle$ to a specific set of final $k$-states $\mid \psi_{f,k}(T)\rangle$. The set of initial $k$-states is related to the transformation of all basis eigenvectors $\{\mid j\rangle\}$ to $\{\mathcal{U}\mid j\rangle\}$, plus the transformation of the initial state $\{\sum_{j=1}^N\mid j\rangle/\sqrt{N}\}$ to $\{\mathcal{U}\{\sum_{j=1}^N\mid j\rangle/\sqrt{N}\}$, which avoids errors due to undesirable relative phases~\cite{universal,Kosloff}.

\subsection{Quantum Permutation Algorithm}
Consider a set with three elements $\{1,2,3\}$. There are six possible permutations for this set, where three have even parity $(1, 2, 3)$, $(3, 1, 2)$, $(2, 3, 1)$; and three have odd parity $(3, 2, 1)$, $(2, 1, 3)$, $(1, 3, 2)$. The objective of the QPA~\cite{Gedik2015} is to determine the parity of the permutations and its protocol can be illustrated by associating the permutation to a function $f(x)$ on the set $x\in \{1,2,3\}$. Classically, one must evaluate $f(x)$ for two different values of $x$, while the QPA can determine the parity with a single evaluation of $f(x)$~\cite{Gedik2015}.

In order to show how the QPA works, we use the qutrit as unit of quantum information to encode the three elements of the set, which can be written as
\begin{equation}\label{base}
| 1 \rangle =\begin{pmatrix}1\\
0\\
0
\end{pmatrix},\;
| 2 \rangle=\begin{pmatrix}0\\
1\\
0	
\end{pmatrix},\;
| 3 \rangle=\begin{pmatrix}0\\
0\\
1
\end{pmatrix}.
\end{equation}

 The determination of the parity of a permutation is equivalent to the determination of the parity of six permutation functions, which we represent with six unitary operators. Three of this operators are associated to the even permutations $\Pi_{1}$, $\Pi_{2}$ and $\Pi_{3}$ that map the set of states $(|1\rangle, |2\rangle, |3\rangle)$ to the set of states $(|1\rangle, |2\rangle, |3\rangle)$, $(|3\rangle, |1\rangle, |2\rangle)$, and $(|2\rangle, |3\rangle, |1\rangle)$, respectively. The even permutations operators are cast as
\begin{equation}\label{even} 
% \begin{aligned}
\Pi_{1}= 
\begin{pmatrix}
1 & 0 & 0\\
0 & 1 & 0\\
0 & 0 & 1
\end{pmatrix}
,
\Pi_{2}=  
\begin{pmatrix}0 & 1 & 0\\
0 & 0 & 1\\
1 & 0 & 0
\end{pmatrix}
,
\Pi_{3}=  
\begin{pmatrix}0 & 0 & 1\\
1 & 0 & 0\\
0 & 1 & 0
\end{pmatrix}.
\end{equation}

The three operators associated to the odd permutations $\Pi_{4}$, $\Pi_{5}$ and $\Pi_{6}$ that respectively map the set of states $(|1\rangle, |2\rangle, |3\rangle)$ to the set of states $(|3\rangle, |2\rangle, |1\rangle)$, $(|2\rangle, |1\rangle, |3\rangle)$, and $(|1\rangle, |3\rangle, |2\rangle)$ are

\begin{equation}\label{odd}
 \Pi_{4}= 
\begin{pmatrix}
0 & 0 & 1\\
0 & 1 & 0\\
1 & 0 & 0
\end{pmatrix}
,
\Pi_{5}= 
\begin{pmatrix}
0 & 1 & 0\\
1 & 0 & 0\\
0 & 0 & 1
\end{pmatrix}
,
\Pi_{6}= 
\begin{pmatrix}
1 & 0 & 0\\
0 & 0 & 1\\
0 & 1 & 0
\end{pmatrix}.
\end{equation}

In order to implement the QPA, the system must be initialized in state $| 2 \rangle$, then the quantum fourier transform (QFT) of a qutrit is applied to this initial state, which is given by
\begin{equation}\label{UFT}
U_{FT}=\frac{1}{\sqrt{3}}\left(
\begin{array}{ccc}               
1&1&1 \\                                                       
1&\exp(i2\pi/3)&\exp(-i2\pi/3) \\
1&\exp(-i2\pi/3)&\exp(i2\pi/3) \\
\end{array}\right).
\end{equation}
The obtained state is the following superposition of three states
\begin{equation}
 \mid\psi_1\rangle=\frac{\mid 1\rangle+\exp(i2\pi/3)\mid2\rangle+\exp(-i2\pi/3)\mid 3 \rangle}{\sqrt{3}}.
\end{equation}

\begin{figure}[h]%[bth]
  \centering%\myfloatalign
 {\includegraphics[width=8.0 cm]{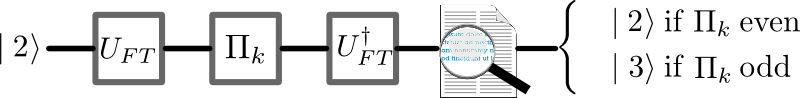}}\\
  \caption{Quantum permutation algorithm (QPA).}
  \label{circuit}
\end{figure}

In Fig.~\ref{circuit} we show a quantum circuit that represents the QPA. The second gate in Fig.~\ref{circuit} encodes one of the six possible permutations operators $\Pi_{k}$, with $k = 1,2,..,6$. The resulting state is $\mid\psi_k\rangle = \Pi_{k} \mid\psi_1\rangle$. To determine the parity of the permutation, one must apply the gate $U_{FT}^\dagger$ and measure the system to check the six possible outcomes, which are described by the following states

\begin{equation} 
\begin{array}{c | c} 
\textbf{Even} & \textbf{Odd} \\ \hline \\ U_{FT}^\dagger \mid \psi_1 \rangle = \mid 2 \rangle & U_{FT}^\dagger \mid \psi_4 \rangle = e^{-2\pi i/3}\mid 3 \rangle \\ U_{FT}^\dagger \mid \psi_2 \rangle = e^{2\pi i/3}\mid 2 \rangle & U_{FT}^\dagger \mid \psi_5 \rangle = e^{2\pi i/3}\mid 3 \rangle\\ U_{FT}^\dagger \mid \psi_3 \rangle = e^{-2\pi i/3}\mid 2 \rangle & U_{FT}^\dagger \mid \psi_6 \rangle = \mid 3 \rangle
\end{array}
\end{equation}

If the measured state is $\mid2\rangle$ ($\mid3\rangle$), the parity of the applied permutation was even (odd). In such a way, only one evaluation of $f(x)$ is necessary in contrast with the classical algorithm, where two evaluations of $f(x)$ are required. An interesting feature of this algorithm is the fact that entanglement is not necessary and only a superposition with well-defined relative phases is necessary. 
 
\section{Results and Analysis}
We use the Hamiltonian of the hybrid qubit system extracted from Ref.~\onlinecite{Shi2014}, which is given by
\begin{eqnarray}
H(\varepsilon)=\begin{bmatrix}
\varepsilon/2 & 0 & \hbar\Delta_1 &   -\hbar\Delta_2   \\[0.3em]
       0 & \varepsilon/2+ \hbar\delta E_L&-\hbar\Delta_3 &\hbar\Delta_4\\[0.3em]
              \hbar\Delta_1 & -\hbar\Delta_3 &-\varepsilon/2 & 0\\[0.3em]
              -\hbar\Delta_2 & \hbar\Delta_4 &0 &-\varepsilon/2 + \hbar\delta E_R
     \end{bmatrix}
\end{eqnarray}
where $\Delta_1/2\pi=2.62$~GHz, $\Delta_2/2\pi=3.5$~GHz, $\Delta_3/2\pi=4.6$~GHz, $\Delta_4/2\pi=1.65$~GHz, $\delta E_L/2\pi=52.7$~GHz, $\delta E_R/2\pi=9.2$~GHz, and $\varepsilon$ is the detuning. This Hamiltonian can be diagonalized for each value of detuning and the resulting energy levels are shown in Fig.(2).
A qutrit requires only three quantum levels, but we take into account four states as our basis of states to account for leakage effects. The ground state is labelled as $|2\rangle$, while the first excited state is labelled as $|1\rangle$. Such a change is due to the QPA, which is initialized in the state $|2 \rangle$ and we choose the initial state as the ground state of the system. The time evolution of the system is given by
\begin{equation}
\frac{\partial|\psi(t)\rangle}{\partial t}=\frac{1}{i\text{\ensuremath{\hbar}}}\left(H_0-\mu E(t)\right)|\psi(t)\rangle,
\end{equation}
where $H_0=Z^{-1}H(\varepsilon_0)Z$ is the diagonalized Hamiltonian for a reference detuning $\varepsilon_0$ that provides a basis for the time evolution of the system. Each column of the $Z$ matrix is given by each eigenvector of $H(\varepsilon_0)$. The dipole type matrix is obtained by $\mu=Z^{-1}H_DZ$, where $H_D$ is a matrix composed by only the diagonal elements of $H(\varepsilon=1)$ that are proportional to $\varepsilon$ and $E(t)=\varepsilon(t)-\varepsilon_0$. The optmized  field can be obtained using the TBQCP method~\cite{control2010}, described in section II (A). 
\begin{figure}[t]%[bth]
        \centering%\myfloatalign
 {\includegraphics[scale=0.57]{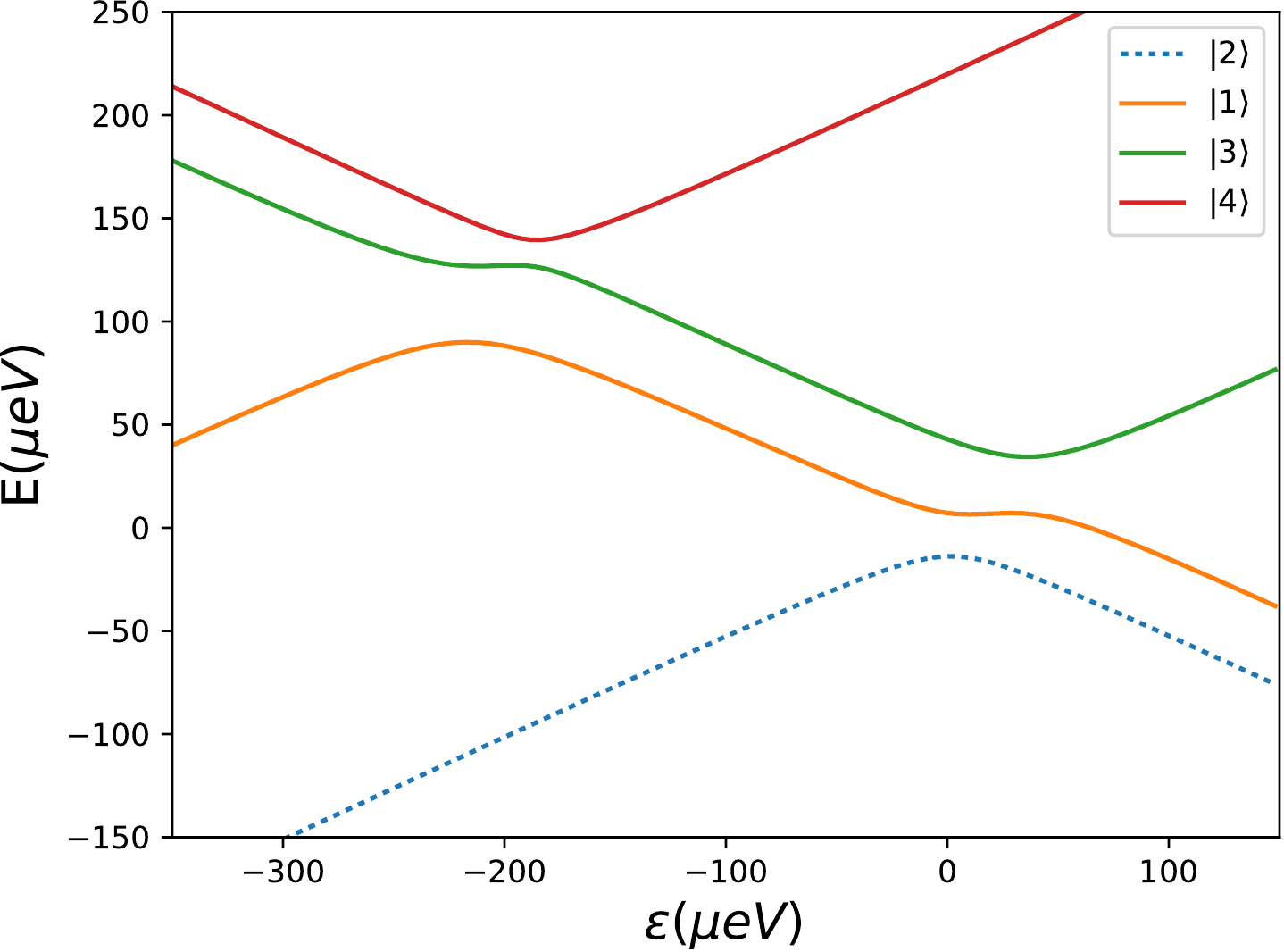}}
  \caption{ The energy levels as a function of the detuning.}
  \label{fig2}
\end{figure}

\begin{figure}[th]%[bth]
        \centering%\myfloatalign
 {\includegraphics[scale=0.6]{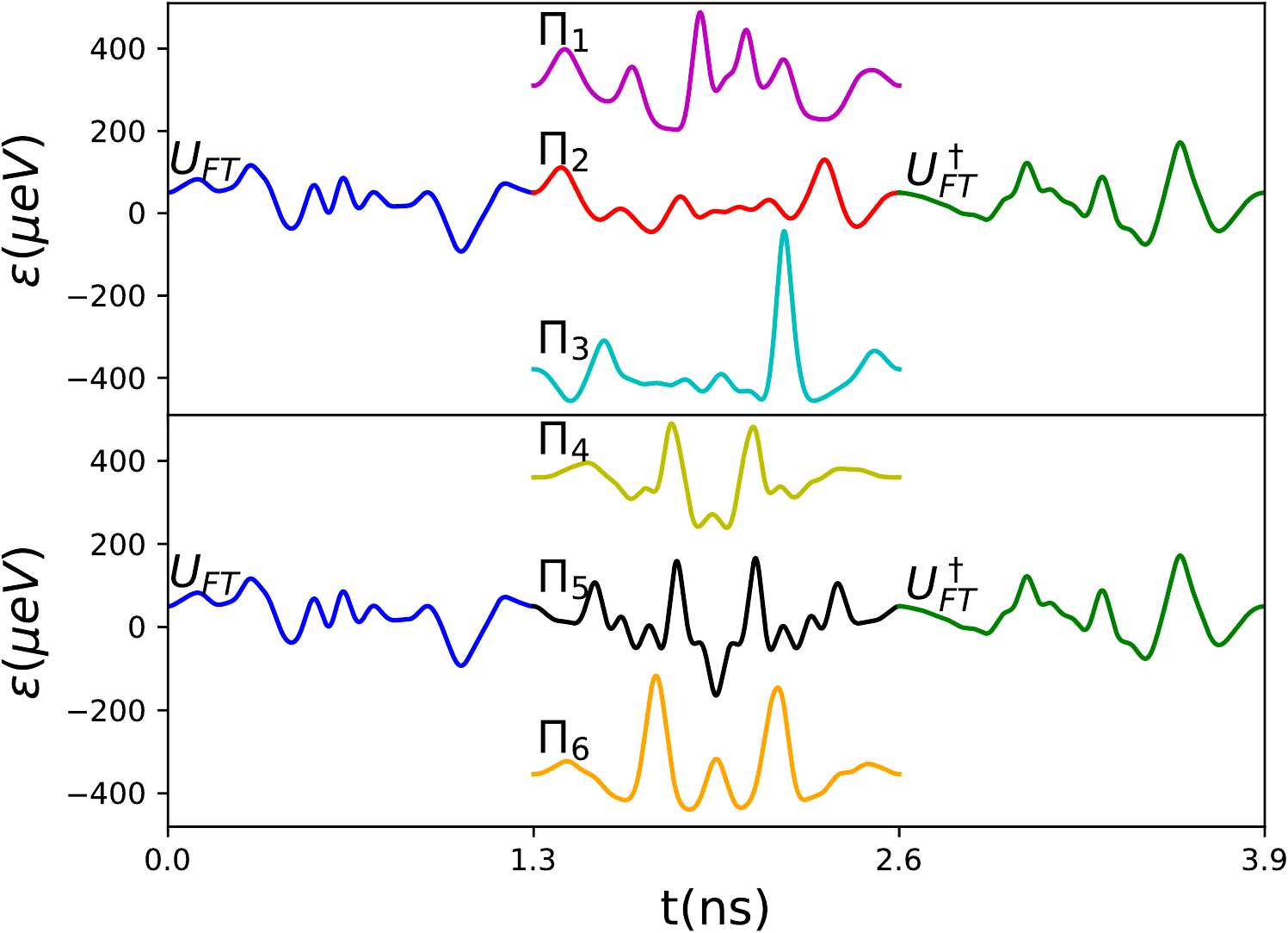}}\\
  \caption{Optimized pulses as a function of time, considering the reference detuning $\varepsilon_0=50$ $\mu$eV. The first gate in each panel corresponds to the $U_{FT}$ gate and the third gate to the $U_{FT}^\dagger$ gate. The six different pulses in the middle correspond to the six permutations gates, three even permutations (top panel) and three odd permutations (bottom panel). The gates $\varPi_1, \varPi_3, \varPi_4,$ and $\varPi_6$ are shown with an offset to better view, but their real reference detuning is $\varepsilon_0=50\mu$eV, such all other gates.}
  \label{PQWpermu}
\end{figure}

Relaxation and decoherence sources are always present and are responsible for reducing coherence and fidelity of a quantum state. We make use of hybrid qubits~\cite{hybrid1,hybrid2}, which presents relaxation time of the order of $20$ ns~\cite{hybrid4}. In our model, we do not take into account relaxation and decoherence effects because we aim at finding control fields that act in a shorter timescale ($T= 1.3$ ns) when compared to the relaxation time.
For the optimization of the control pulses with the TBQCP method, we start with a null trial field. In Fig. \ref{PQWpermu}, we plot the optimized electric fields for all gates required by the QPA and for the reference detuning $\varepsilon_0=50$ $\mu$eV. The first (third) gate corresponds to the $U_{FT}$ ($U_{FT}^\dagger$) gate. The six different pulses in the middle correspond to the six permutations gates, three even (top panel) and three odd (bottom panel).

 \begin{figure}[h]%[bth]
         \centering%\myfloatalign
 {\includegraphics[scale=0.6]{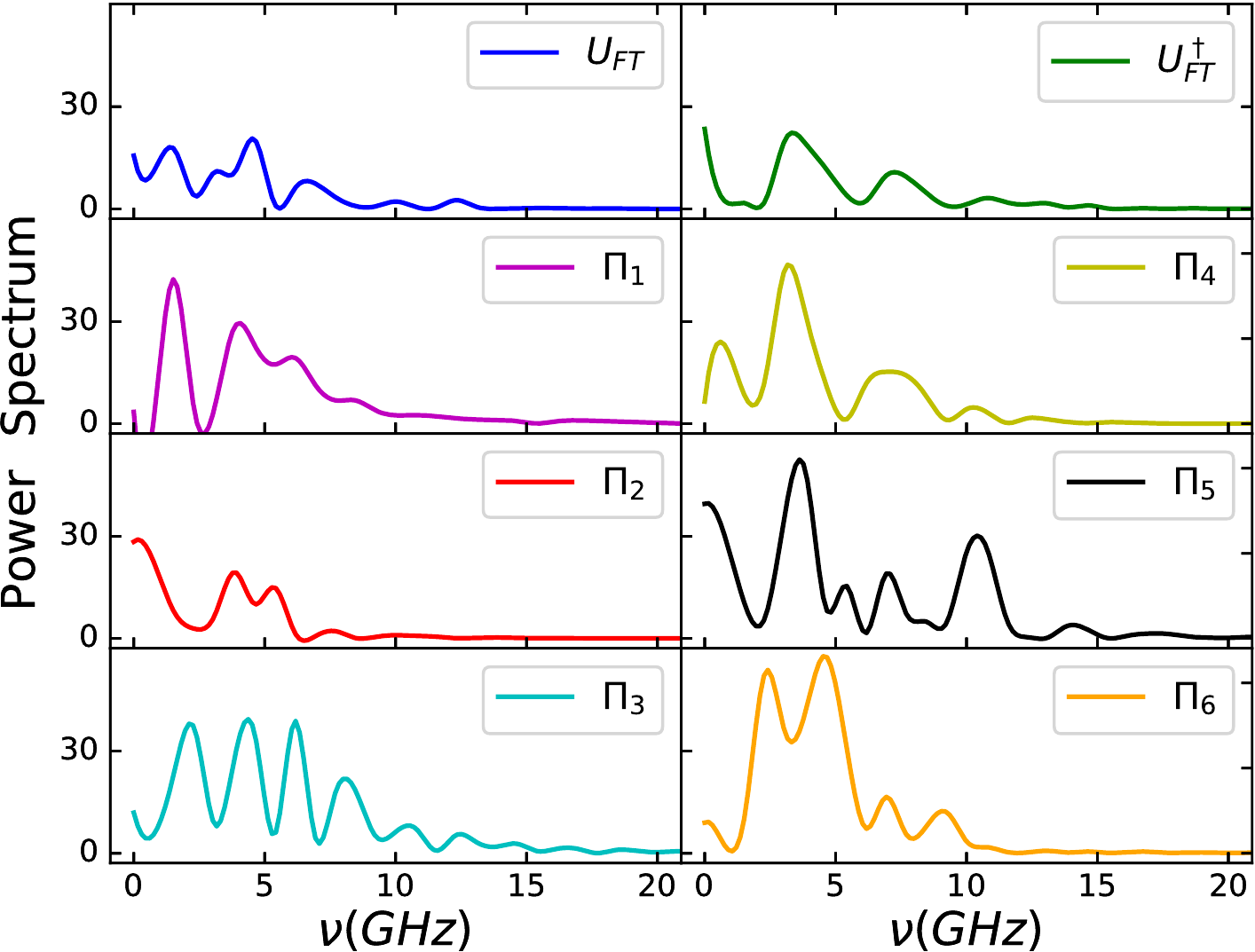}}\\
  \caption{Power spectrum of the electric pulses corresponding to the gates of the QPA.}
  \label{fourier}
 \end{figure}
   
 \begin{figure}[t!]%[htbp]
         \centering%\myfloatalign
 {\includegraphics[scale=0.5]{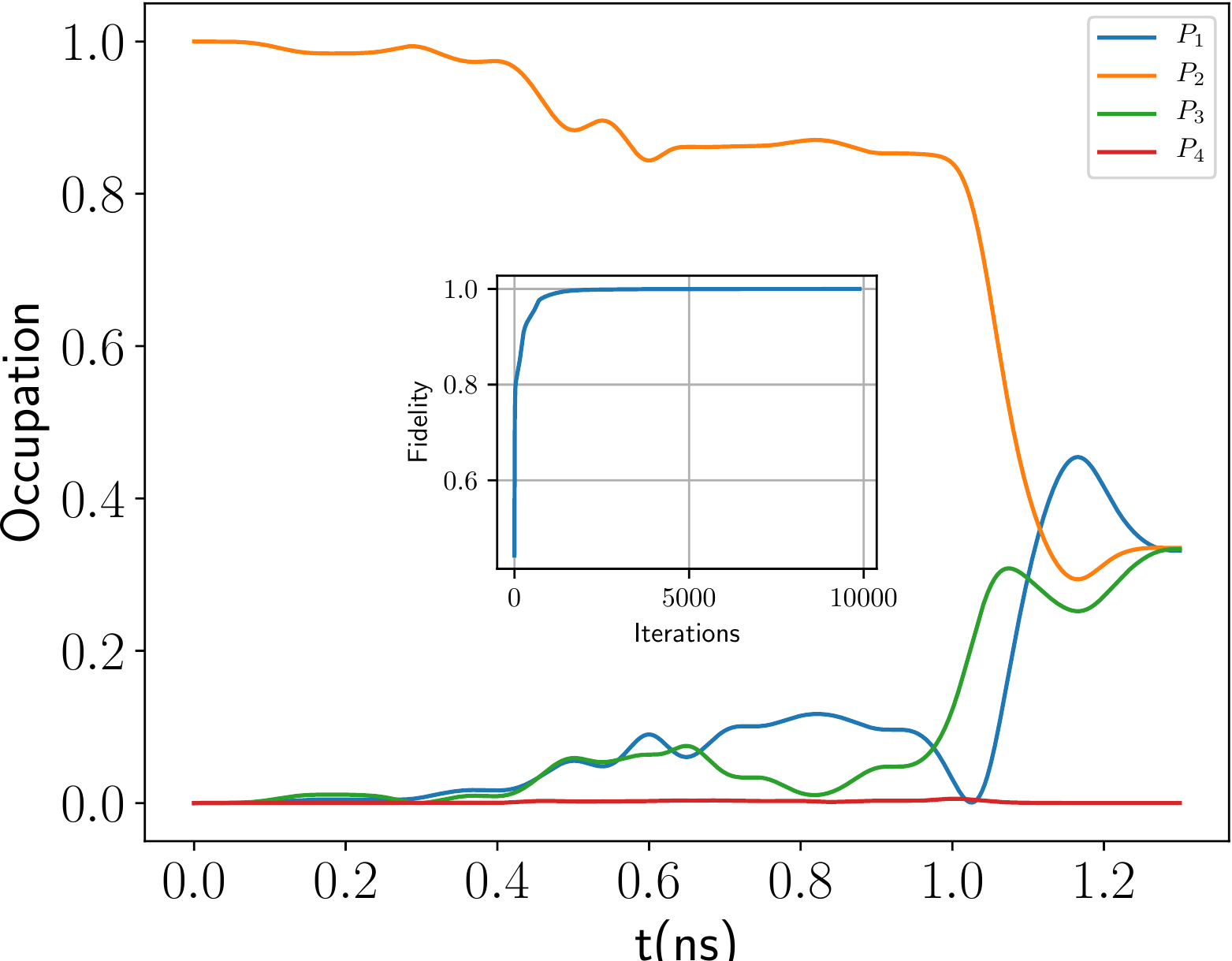}}\\
  \caption{State occupations $P_i=|\langle\psi(t)|i\rangle|^2$ for i=1,2,3,4 as a function of time considering the action of the $U_{FT}$ gate on the initial state $| 2 \rangle$ that converts it to the state $\frac{1}{\sqrt{3}}(\mid 1\rangle+\exp(i2\pi/3)\mid2\rangle+\exp(-i2\pi/3)\mid 3 \rangle)$. The inset shows the fidelity vs. the number of the TBQCP iterations.}
  \label{evolconver}
 \end{figure}

The power spectrum of the optimized pulses (Fig.~\ref{evolconver}) are shown in Fig.~\ref{fourier}. We note that all gates have a nontrivial power spectrum in the range of $0$ and 20 GHz, which is in agreement with the eigenenergies scale of the hybrid qubit (see Fig.~\ref{fig2}).
In Fig.~\ref{evolconver}, we plot the evolution of the state occupation in the four state basis for the reference detuning $\varepsilon_0=50$ $\mu$eV, for the conversion of the initial state $| 2 \rangle$ by the $U_{FT}$ gate into the target state $\mid \psi_1 \rangle$.
%\begin{equation}
%\mid 2 \rangle  \xrightarrow{U_{FT}} \mid \psi_1 \rangle%\frac{1}{\sqrt{3}}(\mid 1 \rangle + \mid 2 \rangle + \mid 3 \rangle) e^{i\phi_1},
%\end{equation}
 In such evolution, we note that there is a very small leakage outside the first three states (the qutrit basis), and the optimal pulse drives the dynamics to yield the target with high fidelity at the final time (see inset o Fig.~\ref{evolconver}). The fidelity is defined as $\mathcal{F}=|\langle\psi(T)|\psi_{target}\rangle|^2$, where $|\psi(T)\rangle$ is the time evolved state and $|\psi_{target}\rangle$ is the desired state under the action of the quantum gate. In the inset o Fig.~\ref{evolconver}, we plot the fidelity for the $U_{FT}$ gate as a function of the number of iterations of the TBQCP. For all gates the leakage to the fourth level was not significant and the fidelity is bigger than 0.9997 for T=1.3 ns. The fidelity depends on the time duration of the pulse, which defines some restrictions for the maximum achieved fidelity. Such dependence is related to the quantum speed limit, which is connected to the minimum time to perform a transition between two states~\cite{qsl1}. Moreover, the connection between quantum speed limit and optimal quantum control has already been investigated~\cite{qsl2,qsl3}. In Ref.~\onlinecite{qsl3}, authors related the quantum speed limit with the infidelity $\mathcal{I}=1-|\langle\psi(T)|\psi_{target}\rangle|^2$ considering the Krotov-algorithm~\cite{krotov}. In table I, we present results for the infidelity $\mathcal{I}$ for all quantum gates required by the QPA considering two different pulse duration (T=1.0 and 1.3 ns). The infidelity in table I must be multiplied by $10^{-5}$ to get its real value, \textit{e.g.}, for the $U_{FT}$ the infidelity is 3.9$\times10^{-5}$ at T=1.3 ns, which corresponds to a fidelity of 0.999961. The worst case shown in table I is for the $\varPi_5$ gate considering a pulse of T=1 ns, which has the infidelity $\mathcal{I}=1414.7\times 10^{-5}$ ($\mathcal{F}=0.985853$) and we attribute such a result to the approach of the quantum speed limit. In general, the infidelity in table I decreases with the increasing of the pulse duration. Only for $\varPi_1$ the infidelity is bigger for T=1.3 ns than for T=1.0 ns. Such a result is due to numerical calculations and the convergence criteria, which was set to stop the TBQCP iterations when the infidelity starts to fluctuate within 20 iterations, \textit{i.e.} achieves a bigger value of infidelity when compared to the infidelity evaluated 20 iterations before.

\begin{table}
\begin{tabular}{|c||c|c|c|c|c|c|c|c|}
\hline 
T (ns) & $ U_{FT}$ & $ \varPi_{1}$ & $ \varPi_{2}$ & $ \varPi_{3}$ & $ \varPi_{4}$ & $ \varPi_{5}$ & $ \varPi_{6}$ & $ U_{FT}^{\dagger}$\tabularnewline
\hline 
\hline 
1.0 &  261.5
 & 1.6 & 90.2 & 916.5 & 65.9 & 1414.7 & 750.4 & 533.6\tabularnewline
%\hline 
%1.25 & 0.2 & 2.3 & 16.9 & 131.3 & 3.9 & 1.5 & 1.5 & 2.9\tabularnewline
\hline 
1.3 & 3.9 & 6.5 & 3.3 & 26.8 & 3.9 & 26.0 & 5.3 & 4.6\tabularnewline
\hline 
\end{tabular}

\protect\caption{Infidelity for all quantum gates necessary to implement the QPA considering different pulse durations (T=1.0,  and 1.3 ns). The infidelity for each case must be multiplied by $10^{-5}$ to get its real value, \textit{e.g.}, for the $U_{FT}$  the infidelity is 3.9$\times10^{-5}$ at T=1.3 ns, which corresponds to a fidelity of 0.999961.
 The number of iterations for each quantum gate is different for each case, ranging from a minimum of 654 iterations for the $U_{FT}$ gate with
$T=1.3$ ns and a maximum of 15425 iterations for the $\varPi_{2}$
gate with $T=1.0$ ns.}
\end{table}

\section{Conclusions}
In this study, we have proposed a physical realization of the QPA by using the platform of hybrid qubit in DQDs.  We employ the TBQCP method to optimize electric pulses that drives the states of the system to the desired set of target states required by QPA. These pulses perform the quantum gates of the QPA with a high fidelity (higher than $0.9997$). Moreover, the short duration of our pulses ($1.3$ ns) compared to the decoherence time ($20$ ns) reported in Ref.~\onlinecite{hybrid4}, support the idea that relaxation would not have significant effects on the performance of our implementation. This fact and our results suggest that the proposed scheme, based on quantum optimal control theory, might be very useful to achieve optimal electric pulses able to manipulate qubits and/or qudits in DQDs. Moreover, we show that the high fidelity achieved in our simulation occurs when the pulse duration is above the quantum speed limit.

\end{document}